\documentclass[aps,prl,twocolumn,superscriptaddress,showpacs]{revtex4}

\usepackage{graphicx}

\begin{document}

%\preprint{}

\title{Evidence of local magnetic order in hcp iron from Raman mode
splitting}


\author{Gerd Steinle-Neumann}
\email[]{g.steinle-neumann@gl.ciw.edu}
\altaffiliation{now at: Geophysical Laboratory, Carnegie Institution of Washington, Washington, DC 20015-1035}
%\thanks{}
\affiliation{Department of Geological Sciences, University of Michigan, Ann Arbor, MI 48109-1063}

%\altaffiliation{now at: Geophysical Laboratory, Carnegie Institution of Washington, Washington, DC 20015-1035}
\author{Lars Stixrude}
\affiliation{Department of Geological Sciences, University of Michigan, Ann Arbor, MI 48109-1063}

\author{R.\ E.\ Cohen}
\affiliation{Geophysical Laboratory, Carnegie Institution of Washington, Washington, DC 20015-1035}

\author{Boris Kiefer}
\affiliation{Department of Geological Sciences, University of Michigan, Ann Arbor, MI 48109-1063}


\date{\today}

\begin{abstract}
Experimental measurements of Raman spectra for hcp iron at high pressure show
two modes over a considerable pressure range in contrast
to the prediction of one doubly degenerate mode for the hcp lattice. We use
density functional theory to investigate the influence of magnetic order on the
Raman active modes of hcp iron. We find an
antiferromagnetic state that lifts the degeneracy of the transverse
optical mode, and yields stable antiferromagnetic moments up to approximately
60 GPa (55 Bohr$^3$).
The resulting frequencies of the two transverse optical modes
are in good agreement with the experimental Raman shifts, lending support
to the existence of local antiferromagnetic order in hcp iron.
\end{abstract}

%\keywords{}

\maketitle

%\section{}
%\subsection{}
%\subsubsection{}

The presence of magnetism strongly influences material properties
and phase relations in the $3d$ transition metals. This is particularily true
for iron where the competition between magnetic and non-magnetic contributions
to the internal energy, differences in the vibrational and magnetic entropy,
and differences in volumes all contribute to phase stability \cite{moroni}.
Ferromagnetism
stabilizes the body centered cubic structure over the close packed phases
at ambient condition \cite{bagno}, and an understanding of the spin-density
wave ground state in its high temperature polymorph, cubic close packed
(fcc), has important implications on its physical properties
\cite{uhl}. Magnetism in the high pressure polymorph, hexagonally
close packed (hcp), is less well characterized and has long been thought to
be absent, but its proposed presence \cite{steinle} may play an important role
in our understanding of material properties at pressure, with
applications to the study of planetary interiors \cite{stixrude} and impact
phenomena.

\begin{figure}[b]
\includegraphics[width=55mm]{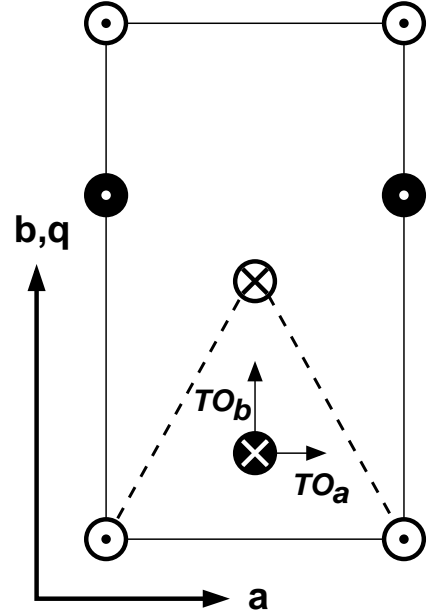}
\caption{\label{afm}
Orthorhombic antiferromagnetic groundstate of hcp iron (afmII).
Open circles show the atomic positions at $z$=1/4, filled circles at
$z$=3/4 with the arrows indicating the direction of spin on the atoms.
The orthorhombic unit cell is outlined with the axes given.
The direction of the wavevector $q$ for the afmII spinwave is along the
$b$-axis. The $a$- and $b$-axes also define the eigenvectors for the $TO$
modes ($TO_a$ and $TO_b$). The $c$-axis is out of the plane.}
\end{figure}

The hcp lattice has two atoms per unit cell giving rise to optical phonon
modes through Brillouin zone folding. Advances in experimental techniques 
have made it possible to measure the Raman active transverse optical ($TO$) 
mode in iron through optical spectroscopy \cite{merkel,olijnyk}.
In contrast to the fundamental prediction that this mode ($E_{2g}$) be doubly
degenerate, one set of experiments at high pressure show
two peaks in the Raman spectrum up to pressures of 40 GPa \cite{merkel}. 
While amplitude and sharpness of the two peaks differ significantly, this 
observation nonetheless suggests that the symmetry of hcp iron 
is lower than the atomic arrangement; spin-phonon interactions provide a
symmetry-breaking mechanism.  Based on first principles
theory the lowest energy state found to date is an anti-ferromagnetic (afm)
structure with orthorhombic symmetry 
and four atoms in the unit cell (afmII) \cite{steinle}.
The afmII structure corresponds to a spin wave with wave vector $q=(0,1/2,0)$,
at the $M$ point on the Brillouin zone boundary (Fig.\ \ref{afm}).
This structure has been predicted to be stable for hcp iron
up to pressures of almost 60 GPa \cite{steinle}. 
The magnetic structure results in two $TO$ zone center modes ($A_{2g}$) both
of which are Raman active.

Here we investigate the influence of afmII ordering on the
zone center $TO$ phonon mode over the compression range where finite moments arepredicted ($\ge$60 Bohr$^3$/atom). We calculate the zone center Raman
frequencies by the frozen phonon approximation:
energy changes are evaluated in response to small displacements along the
phonon eigenvector, the second order term yielding the
frequency. We base our analysis on total energies obtained with the
full-potential linearized-augmented plane-wave method (LAPW) \cite{singh}
using a generalized gradient approximation \cite{pbe} to the exchange
correlation potential. We treat $3s$, $3p$, $3d$, $4s$, and $4p$ states as
valence electrons for all volumes and use
$R_{MT}$=2.0 Bohr for the muffin tin radii, $R_{MT}K_{max}=9.0$,
a 12$\times$6$\times$12 special k-point mesh, and a temperature
broadening of  5 mRy. With this set
of computational parameters we have previously established convergence of
relative energies to within $0.1$ mRy/atom and magnetic moments to better
than $0.05$ $\mu_{B}$/atom \cite{steinle}. The $TO$ modes are
characterized by displacements of the close packed planes with respect to one
another. The eigenvectors for the the two $A_{2g}$ modes in the afmII
structure are along the orthorhombic $a$- and $b$-axes (Fig.\ \ref{afm}).
For comparison we also compute the $TO$ mode frequency for the non-magnetic
case. To efficiently evaluate the energetics of the system we fixed the axial
ratio in our calculations to $c/a=1.6$, close to the experimentally  
\cite{jephcoat,mao90} and theoretically \cite{steinle} estimated equilibrium
value. 

\begin{figure}[b]
\includegraphics[width=85mm]{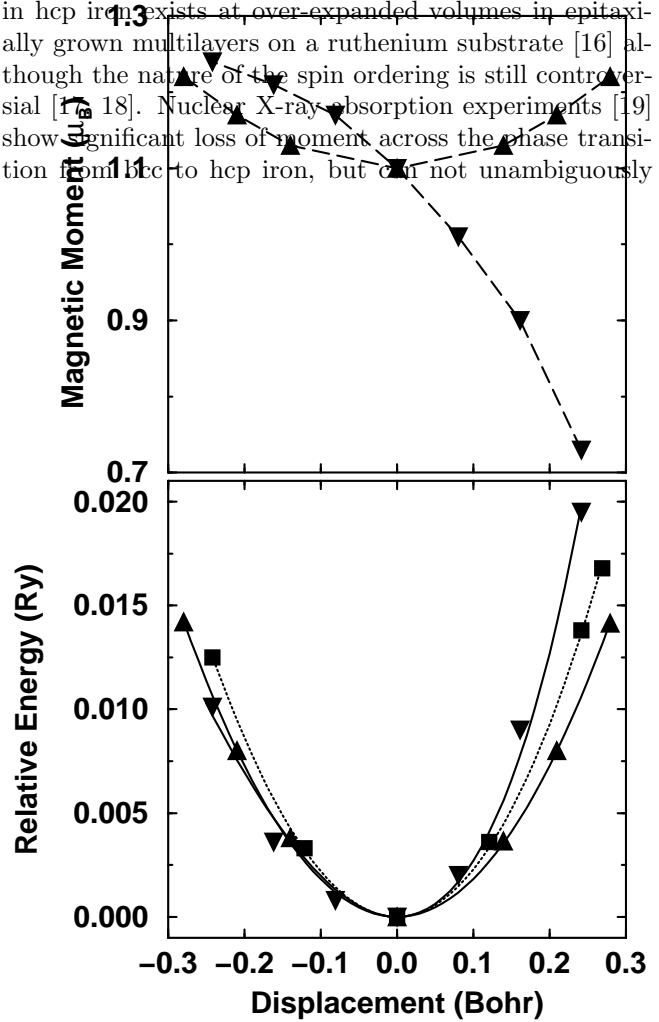}
\caption{\label{displ}
Energy of displacement of hcp iron at V=70 Bohr$^3$/atom in the lower
panel.  Relative energy changes as a function of displacement
for the non-magnetic structure (squares),
and the afmII structure for displacements along $TO_a$ (triangles up)
and $TO_b$ (triangles down). Dotted and solid lines are third order
polynomial fits to the results. The upper panel shows the change in the afmII
moment as a function of displacement for $TO_a$ and $TO_b$ (same symbols, dashedline to guide the eye).}
\end{figure}

Inspection of the afmII structure reveals that fundamentally different spin
interactions are involved in the two $TO$ modes (Fig.\ \ref{afm}). 
For displacements along the orthorhombic 
$a$-axis ($TO_a$) atoms approach nearest neighbors with unlike spin.
For displacements along the 
orthorhombic $b$-axis ($TO_b$) atoms alternately move
towards a nearest neighbor with like spin 
and towards a pair of nearest neighbors of opposite spin.
The resulting energy - displacement relations reflect the antiferromagnetic
spin interactions (Fig.\ \ref{displ}): for unlike spins approaching the energy
is reduced with respect to the non-magnetic case, and for same spin nearest
neighbor interactions (along positive $b$) there is an additional repulsion.
The calculated equilibrium magnetic moments are consistent with these
findings (Fig.\ \ref{displ}): They increase for displacements along $a$,
as they do along negative $b$; for positive $b$ they decrease considerably as
the like spin nearest neighbors approach.

\begin{figure}
\includegraphics[width=85mm]{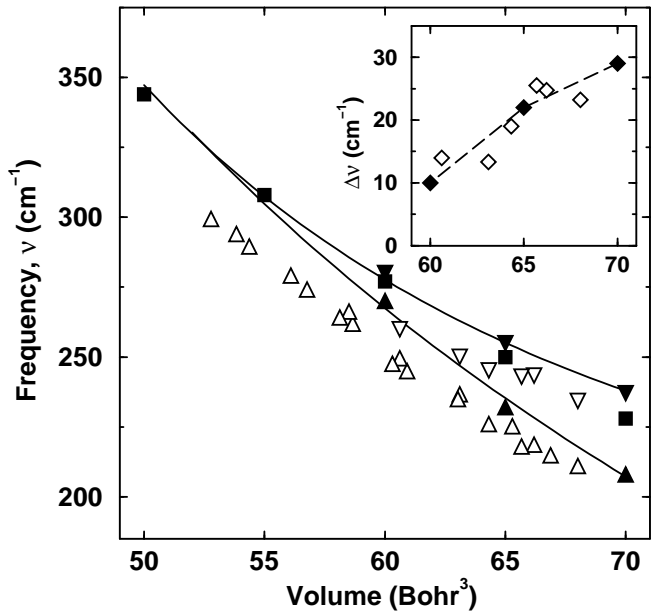}
\caption{\label{raman}
Raman frequencies as a function of atomic volume.
Non-magnetic calculations are shown in filled squares. The afmII structure
results in transverse optical frequencies (filled triangles) with $TO_a$
being the lower (up) and $TO_b$ (down) the upper branch. The solid lines
through $TO_a$ and $TO_b$ are third order polynomial fits in $V^{-2/3}$.
Experiments (Ref.\ \protect\onlinecite{merkel}) identify two peaks in the
Raman spectra up to 40 GPa in open symbols. The stronger peak is shown in the
triangles up, and the weaker peak in triangles down. The inset compares the
split in Raman frequencies from theory (filled diamonds) and experiment
(open diamond).}
\end{figure}

We find that the $TO_a$ mode frequency agrees well with that of the lower
frequency, higher amplitude peak found in the Raman experiments, and that
the $TO_b$ mode frequency corresponds to the experimentally observed  
satellite peak at higher frequency (Fig.\ \ref{raman}). The magnitude of the
predicted $TO$ mode splitting decreases as the afm moment is reduced by
compression, in excellent agreement with the observation in the Raman
experiments (Fig.\ \ref{raman}).
The magnitude of splitting is related to the directionality of the 
spin wave considered here (with wave vector $M$), but other spin waves
with wavevectors along high symmetry directions in the base of the hexagonal
Brillouin zone will generally result in splitting of the $TO$ mode.
The systematic offset of the calculated frequencies by approximately 20
cm$^{-1}$ ($<$ 10\%) is typical for a comparison of computed and measured
phonon frequencies \cite{znolijnyk,althoff}.

Spin-phonon interactions have been found to have a strong
effect on Raman scattering in a number of systems, including cupric oxide
\cite{chen95} and the copper-ruthenium oxide RuSr$_2$GdCu$_2$O$_8$
\cite{fainstein}.  The general character of the effect of spin-phonon
interaction on the Raman spectra in these materials is consistent with the
observations for hcp iron: broad, low amplitude, satellite peaks appear as 
the sample is cooled below the Curie or Neel temperature, upon further cooling
such peaks generally sharpen and gain amplitude as the degree of magnetic 
ordering increases. Observation of an increase in amplitude of the satellite
peak in hcp iron upon cooling would lend support to the predicted magnetic
structure.

Other experimental investigations of possible magnetic states in hcp
iron have been inconclusive. 
Magnetism in hcp iron exists at over-expanded volumes in epitaxially
grown multilayers on a ruthenium substrate \cite{maurer} although the nature
of the spin ordering is still controversial \cite{lager,knab}.
Nuclear X-ray absorption experiments \cite{rueff}
show significant loss of moment across the phase transition from bcc to
hcp iron, but can not unambiguously be interpreted to show no
moments in the high pressure polymorph: the change in absorption spectra
is due to changes in the density of states as well as to spin related
satellites.
Diamond anvil cell in-situ M\"ossbauer measurements on hcp iron have not
shown any ordered magnetic ground state in its
stability field \cite{nicol,cort,taylor1,taylor2} but it was recognized that
magnetism cannot unambigously be ruled out based on the data
\cite{nicol,taylor1}.
The M\"ossbauer data suggest that correlation times of spin fluctuations must
be relatively short, or that competing contributions to the effective
field cancel each other as is the case in CaRuO$_3$
perovskite \cite{gibb,felner}.
The in-situ observation of super-paramagnetism in hcp-Fe \cite{taylor1} may
support the notion that the effective field is approximately cancelled, and
suggests, following the procedure of Felner {\it et al.} \cite{felner}, a
Mossbauer experiment on an appropriately doped hcp-Fe sample (e.g. with Ru).
The inherent frustration of the triangular lattice might lead one to
suspect more complex spin arrangments than those considered here, such as
incomensurate spin waves, as in the case of fcc iron \cite{tsunoda,uhl},
non-colinear structures, spin glass, or a combination of these \cite{cohen01}.

\begin{table}[t]
\caption{\label{nutab}
Raman shift for hcp iron for the non-magnetic $E_{2g}$ mode and
the two antiferromagnetic $TO$ modes ($A_{2g}$: $TO_a$, $TO_b$) as a function
of atomic volume. Pressure $P$ is from the afmII equation of state
(Ref.\ \protect\onlinecite{steinle}).}
\begin{ruledtabular}
\begin{tabular}{ccccc}
Volume     & $P$   & $\nu$ ($E_{2g}$)& $\nu$ ($TO_a$) & $\nu$ ($TO_b$)\\
(Bohr$^3$) & (GPa) &(cm$^{-1}$)      &(cm$^{-1}$)     &(cm$^{-1}$)    \\
\colrule
70         & 2     & 228        & 208              & 237              \\
65         & 22    & 250        & 232              & 254              \\
60         & 55    & 277        & 270              & 280              \\
55         & 109   & 308        & --               & --               \\
50         & 193   & 344        & --               & --               \\
\end{tabular}
\end{ruledtabular}
\end{table}

Acoustic modes can also be influenced by spin-phonon interactions as is
evidenced by anomalous phonon dispersion in fcc iron near the zone center
\cite{zaretsky}. Indeed we find that the afmII ordering also influences the
compressional properties of hcp iron \cite{steinle}: the
bulk modulus at pressure is reduced, bringing it in closer agreement
with experimental data \cite{jephcoat,mao90} than non spin-polarized
calculations. Better agreement of aggregate elastic properties between
afmII and experimental data lends further, independent, support to the presence
of magnetism in hcp iron.

The possible existence of magnetic correlations in hcp iron may be important
in our understanding of the recent observation of superconductivity in hcp 
iron under pressure \cite{shimizu}. Long thought to be antithetical,
superconductivity and magnetism have recently been observed simultanously
in a number of systems \cite{saxena,pfleiderer,aoki}. Of particular interest
is the observation for ZrZn$_2$ where
magnetism and superconductivity appear to be directly coupled \cite{pfleiderer}
as is evidenced by the loss of superconductivity and magnetism at the same
pressure, leading to the speculation that hcp iron might behave similarly
\cite{saxena01}.  

In conclusion, we report first-principles results on the zone center 
$TO$ phonon frequency of hcp iron for non-magnetic and afm 
structures at ambient and high pressure from a frozen phonon calculation. 
We find good agreement with experimental results for the absolute frequencies,
and spin-phonon interactions provide a quantitative explanation for 
the split of the $TO$ mode in hcp iron which is observed up to a pressure of
40 GPa. 
In combination with a considerably better agreement of elastic properties 
between experiment and the orthorhombic afm structure, as compared to
non-magnetic calculations, this
suggests that magnetism may play an important role in the 
physical behavior of hcp iron at high pressure.

\begin{acknowledgments}
We greatly appreciate helpful discussion with A.\ Goncharov, R.\ Hemley, 
I.\ Mazin, S.\ Merkel, H.\ Olijnyk, and V.\ Struzhkin,
and thank H.\ Krakauer and D.\ Singh for use of their LAPW code. This work was
supported by the National Science Foundation under grants
EAR-9980553 (LS) and EAR-998002 (REC), and by DOE ASCI/ASAP subcontract B341492
to Caltech DOE W-7405-ENG-48 (REC). Computations were performed on the Cray
SV1 at the Geophysical Laboratory, supported by NSF grant EAR-9975753 and by
the W.\ M.\ Keck Foundation.
\end{acknowledgments}

%\bibliography{ms.bib}

\end{document}